\documentclass[aps,prl,showpacs,amsmath,amssymb,amsfonts,twocolumn,
superscriptaddress,floatfix,footinbib]{revtex4}

\usepackage{subfigure}
\usepackage{graphicx}
\usepackage[flushmargin]{footmisc}
\usepackage{color}
\usepackage{soul}
\usepackage{longtable}
\usepackage[normalem]{ulem}

\begin{document}

\title{Spin rings in bi-stable planar semiconductor microcavities}

  \author{C. Adrados}
    \affiliation{Laboratoire Kastler Brossel, Universit\'{e} Pierre et Marie Curie-Paris 6,
    \'{E}cole Normale Sup\'{e}rieure et CNRS, UPMC Case 74, 4 place Jussieu,
    75005 Paris, France}

  \author{A. Amo}
    \affiliation{Laboratoire Kastler Brossel, Universit\'{e} Pierre et Marie Curie-Paris 6,
    \'{E}cole Normale Sup\'{e}rieure et CNRS, UPMC Case 74, 4 place Jussieu,
    75005 Paris, France}

  \author{T. C. H. Liew}
    \affiliation{Institute of Theoretical Physics, \'{E}cole Polytechnique F\'{e}d\'{e}rale de Lausanne, CH-1015, Lausanne, Switzerland}
  \author{R. Hivet}
    \affiliation{Laboratoire Kastler Brossel, Universit\'{e} Pierre et Marie Curie-Paris 6,
    \'{E}cole Normale Sup\'{e}rieure et CNRS, UPMC Case 74, 4 place Jussieu,
    75005 Paris, France}
  \author{R. Houdr\'{e}}
    \affiliation{Institut de Physique de la Mati\`{e}re Condens\'{e}e,
    Facult\'{e} des Sciences de Base, b\^{a}timent de Physique,
    Station 3, EPFL, CH 1015 Lausanne, Switzerland}
  \author{E. Giacobino}
    \affiliation{Laboratoire Kastler Brossel, Universit\'{e} Pierre et Marie Curie-Paris 6,
    \'{E}cole Normale Sup\'{e}rieure et CNRS, UPMC Case 74, 4 place Jussieu,
    75005 Paris, France}
  \author{A. V. Kavokin}
    \affiliation{Dipartimento di Fisica, Universita di Roma II, 1 via Ricerca Scientifica, 00133, Roma, Italy}
    \affiliation{Physics and Astronomy School, University of Southampton, Highfield, Southampton, SO171BJ, UK}
  \author{A. Bramati}
    \affiliation{Laboratoire Kastler Brossel, Universit\'{e} Pierre et Marie Curie-Paris 6,
    \'{E}cole Normale Sup\'{e}rieure et CNRS, UPMC Case 74, 4 place Jussieu,
    75005 Paris, France}


\date{\today}

\begin{abstract}
A unique feature of exciton-polaritons, inherited from their mixed light-matter origin, is the strongly spin-dependent polariton-polariton interaction, which has been predicted to result in the formation of spin rings in real space [Shelykh \emph{et al.}, \emph{Phys. Rev. Lett.} {\bf 100}, 116401 (2008)]. Here we experimentally demonstrate the spin bi-stability of exciton-polaritons in an InGaAs-based semiconductor microcavity under resonant optical pumping. We observe the formation of spin rings whose size can be finely controlled in a spatial scale down to the micrometer range, much smaller than the spot size. We additionally evaluate the sign and magnitude of the antiparallel polariton spin interaction constant.
\end{abstract}

\maketitle
Spin-dependent particle-particle interactions can be found in a wide variety of systems, from quarks~\cite{Eichten1981} and ferromagnets, to spinor atomic Bose-Einstein condensates~\cite{Stenger1998}, and can give rise to multi-stability~\cite{Gippius2007,Paraiso2010}: the possibility of having more than two stable states for a given set of conditions in the material. Spin-dependent interactions are of particular interest in semiconductors, due to their high integrability capabilities, microstructuring, and easy electric and optical control~\cite{Zutic2004}. However, 
 they are very weak due to the dominant role of the direct Coulomb interaction between carriers, and they are only evidenced when the system is taken to the extreme situation of few carrier confinement~\cite{Ono2002} or very high optical excitation density, yet with a very weak response~\cite{Vina1996,Nemec2005}. In this respect, the use of semiconductor microcavities opens a whole new universe in the solid state.

Polaritons are the eigentates in semiconductor microcavities, arising from the strong coupling between quantum well excitons and confined photon modes. They have two possible spin projections on the structure's axis ($s_{z}=+1$ and $-1$), and can be excited, and detected, via circularly polarized photons ($\sigma^{+}$ and $\sigma^{-}$, respectively). 

The main signature of the strong coupling in microcavities is the appearance of a normal-mode (Rabi) splitting between exciton polariton modes~\cite{Weisbuch1992}. This splitting gives rise to a strong asymmetry in the interaction strength between polaritons of the same and opposite spin~\cite{Wouters2007c,Glazov2009,Ostatnicky2009}. Polaritons lying at the bottom of the dispersion curve, with parallel spins, strongly repel each other due to the Coulomb exchange interaction between electrons and between holes (interaction strength $\alpha_{1}$). On the other hand, in the antiparallel spin configuration the exchange interaction would result in intermediate states in which the total spin of each exciton is $\pm2$, thus being uncoupled to the photon modes, and lying at the exciton level located at a quite different energy to that of the considered polaritons. 
For this reason, this interaction process is strongly inhibited, resulting in a reduced polariton-polariton interaction strength $\alpha_{2}$ when their spins are anti-parallel ($\alpha_{1}>>\left|\alpha_{2}\right|$)~\cite{Renucci2005,Paraiso2010} which, additionally, should be attractive according to second order perturbation theory~\cite{Wouters2007c,Glazov2009}. This mechanism can be enhanced and modified by bi-excitonic effects~\cite{Wouters2007c}. In this work, we take advantage of the strongly spin-dependent polariton-polariton interactions, in the regime of spin-bistability~\cite{Baas2004b,Gippius2007}, to control the polariton spin state on a spatial scale much smaller than the size of the optical excitation spot. Following the recent proposal by Shelykh et al.~\cite{Shelykh2008b} we demonstrate the creation of "spin rings" within the excitation spot, whose size can be finely tuned by the intensity and degree of circular polarization of the excitation beam. Additionally, we demonstrate non-linear interactions between polariton populations of opposite spin, a consequence of the non-vanishing value of $\alpha_{2}$, whose magnitude and sign we precisely determine for our experimental conditions. All our results are well reproduced by the solution of the spin-dependent Gross-Pitaevskii equation of the polariton system~\cite{Shelykh2006}\footnote{See supplementary information for the parameters employed for our spin-dependent Gross-Pitaevskii model}.

    \begin{figure}[t]
        \centering
        \includegraphics[width=1\columnwidth]{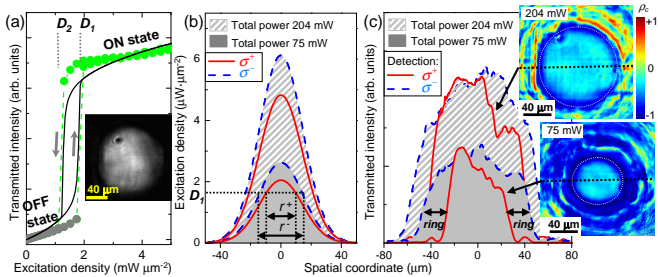}
    \caption{(Color online) (a) Measured (dots) and computed (solid lines) transmission dependence at the center of the spot as a function of excitation density for a purely $\sigma^{+}$ beam. The inset shows a real space image in the ON state. (b) Computed cross section passing through the center of Gaussian spot for the $\sigma^{+}$ and $\sigma^{-}$ components of an elliptically polarized beam ($\theta=-0.2\pi$~rad) at two different powers. 
    (c) Experimental profiles of the transmitted intensity, for two different excitation densities, resulting in spin rings of different radii as evidenced in the spatially resolved degree of polarization of the transmitted light shown in the inset (rings are signaled in white dots). The dashed lines show the coordinate from where the profiles in (d) were obtained.}
    \label{figure1}
    \end{figure}

Our experiments are performed at 6~K in an InGaAs/GaAs/AlGaAs planar microcavity with a Rabi splitting of 5.1~meV, and a polariton linewidth of 0.1~meV at zero exciton-cavity detuning~\cite{Houdre2000}. The excitation is performed at normal incidence with a polarized beam of controlled ellipticity coming from a cw single-mode Ti:Sapphire laser, in a Gaussian spot of 38~$\mu$m in diameter. The detection is polarization resolved and performed in transmission geometry~\cite{Amo2008}.

When the sample is excited with $\sigma^{+}$ polarized light with photon energy 0.124~meV blue-detuned from the lower polariton branch (LPB), we observe a step like behavior in the transmitted intensity versus the excitation density, with a hysteresis cycle [Fig.~\ref{figure1}(a)] giving rise to a bi-stable region~\cite{Baas2004b}, between excitation densities $D_{1}$ and $D_{2}$. This non-linear transmission arises from the renormalization of the dispersion curve when the polariton density is increased via the excitation power. When the system is in the OFF state (lower bi-stable branch), the transmitted intensity is low as the excitation is out of resonance. However, above a given threshold, polariton-polariton interactions give rise to the energy renormalization of the system, and the $\sigma^{+}$-LPB enters in resonance with the excitation laser, resulting in a high transmission (ON state). An analogous curve would be obtained for $\sigma^{-}$ polarized excitation.

The data displayed in Fig.~\ref{figure1}(b) correspond to the transmission in the center of the spot under purely $\sigma^{+}$ polarized excitation. Let's now consider a spot with a Gaussian spatial profile and elliptically polarized excitation. We can divide the excitation spot into two circularly polarized components: $\sigma^{-}$, in a larger proportion, and $\sigma^{+}$, as depicted in Fig.~\ref{figure1}(b). We observe that the $D_{1}$ threshold density between the OFF and ON states is not reached in the whole spot at the same time: $D_{1}$ is reached at a radius $r^{-}$ for the $\sigma^{-}$ profile, which is bigger than the radius $r^{+}$ for the $\sigma^{+}$ profile at the same density. Within these radii the transmission in each polarization is in the ON state, with a large density, while outside it is very low (OFF state). This is what is observed in Fig.~\ref{figure1}(c) for an ellipticity of excitation $\theta$ (phaseshift between the $x$ and $y$ linearly polarized components of the incident light) equal to -0.2$\pi$~rad ($70\%$ $\sigma^{-}$, $30\%$ $\sigma^{+}$), and two different excitation powers. This effect, which arises from the Gaussian distribution of our excitation spot and from the sharp transition from the OFF to the ON state for each polarisation, leads to the appearance of a ring in the spatial profile of the transmitted beam with a degree of circular polarisation $\rho_{c}$ close to 1 [$\rho_{c}=(I^{+}-I^{-})/(I^{+}+I^{-})$, $I^{+(-)}$ being the $\sigma^{+}$ ($\sigma^{-}$) transmitted intensity], as shown in the insets of Fig.~\ref{figure1}(c). The spin ring, whose size is delimited by $r^{+}$ and $r^{-}$, is a domain in which the majority of polaritons have the same spin, corresponding to $\sigma^{-}$ polarisation of emission.

The radius and thickness of the spin ring can be modified by changing, respectively, the total power or the ellipticity of the excitation beam. In the first case, once the centre of the spot overcomes the threshold density $D_{1}$ with the minority polarization, a spin ring is formed. As the excitation power is increased, both $r^{+}$ and $r^{-}$ increase in size, 
 but the thickness of the ring ($r^{+}$ and $r^{-}$) does not change noticeably, as evidenced in the insets of Fig.~\ref{figure1}(c).

    \begin{figure*}[t]
        \centering
        \includegraphics[width=1\textwidth]{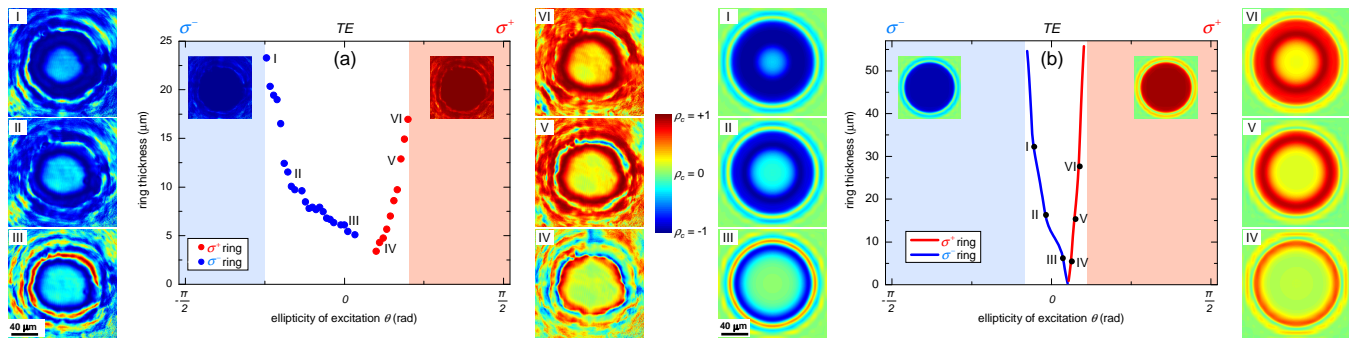}
    \caption{(Color online) Thickness of the spin ring \emph{vs} ellipticity of the excitation beam ($\theta$) as obtained (a) experimentally and (b) from the simulations. The coloured regions indicate the ellipticities for which the whole spot is purely $\sigma^{+}$ or $\sigma^{+}$ (see inset). Surrounding images show the transmitted degree of circular polarization, spatially resolved, for selected values of $\theta$. Red, blue and green correspond, respectively, to purely $\sigma^{+}$, purely $\sigma^{-}$ and linear polarizations. Note that additionally to the main internal spin ring, on which we have performed all our analysis, additional rings are evidenced in the external part of the spot, where the intensity is very low.}
    \label{figure2}
    \end{figure*}

Fine control on the thickness of the spin rings can be obtained by changing the ellipticity of the excitation beam, as the ratio between the $\sigma^{+}$ and $\sigma^{-}$ excitation components changes. When the polarization of excitation is close to linear ($\theta$ close to 0), the ratio between the $\sigma^{+}$ and $\sigma^{-}$ components is almost 1, resulting in similar radii and, consequently, very narrow rings. On the contrary, when the ellipticity of excitation approaches $\pm\pi/2$, the excitation beam is almost purely circularly polarized, leading to a big difference between the radii of both circularly polarized components: the spin rings are wide. This is what is observed in Fig.~\ref{figure2}(a), which shows the thickness of the spin rings obtained as a function of the ellipticity of the excitation beam for a power of 77.4~mW. Real-space images of $\rho_{c}$ corresponding to selected ellipticities are also shown in Fig.~\ref{figure2}(a), panels I-VI. In the middle of the spot, $\rho_{c}$ is almost zero, and it reaches very high values in the ring region. In the external part of the spot, where the transmitted intensity is very low, additional rings are observed. When the ellipticity approaches \emph{zero} we observe a shrinkage of ring thickness, with a minimum value of 3~$\mu$m, more than one order of magnitude smaller than the spot size.

The fact that the minimum ring size is not obtained for strictly linearly polarized excitation, and the asymmetry observed in Fig.~\ref{figure2} around $\theta=0$, arise from the presence of an intrinsic weak polarization splitting in our sample, which slightly rotates the pseudospin of injected polaritons~\cite{Amo2009c}. Simulations based on the solution of the spin dependent Gross-Pitaevskii equation~[27], and accounting for this intrinsic splitting, quantitatively reproduce our observations, as Fig.~\ref{figure2}(b) shows.

\begin{figure}[t]
        \centering
        \includegraphics[width=1\columnwidth]{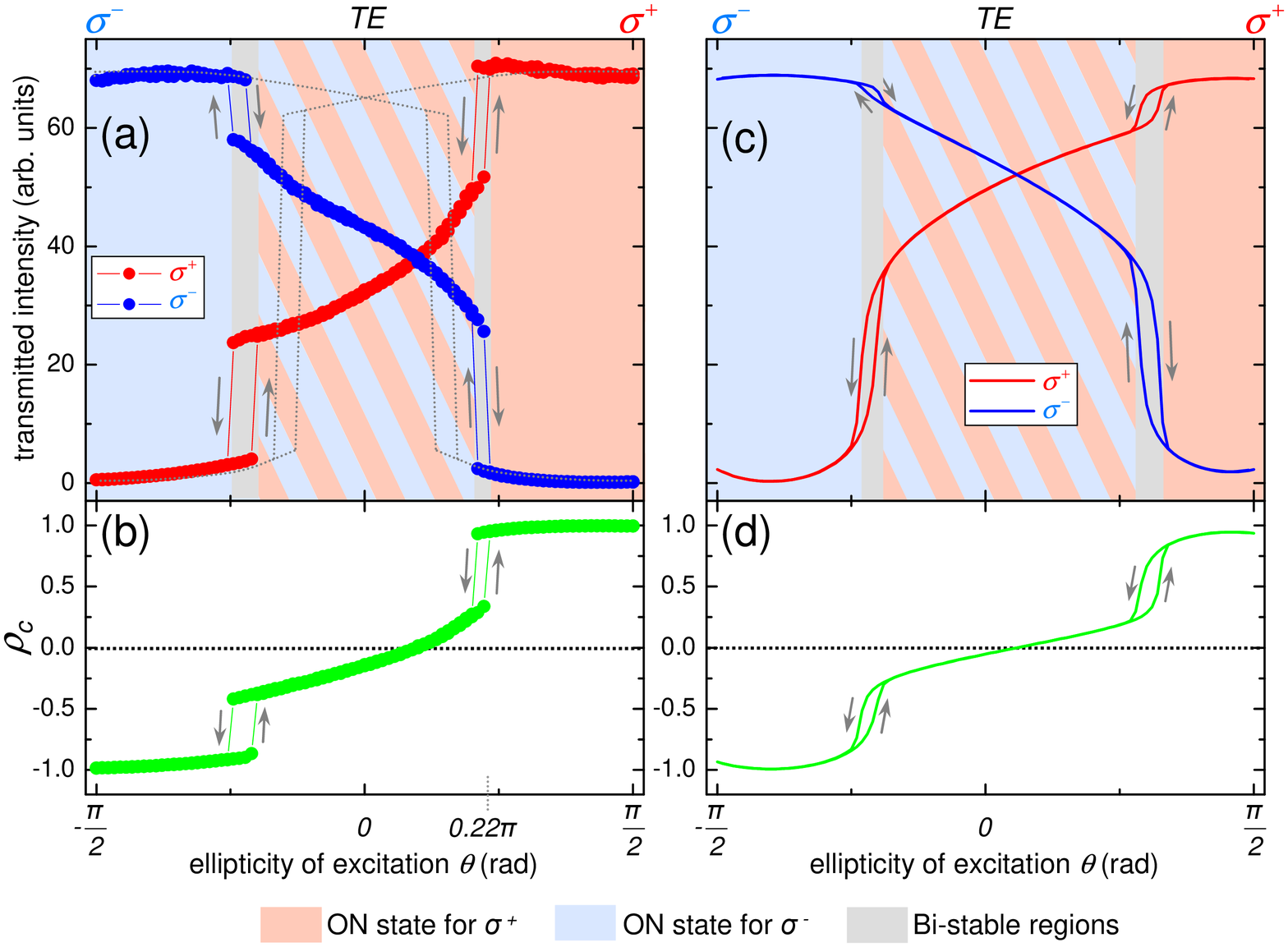}
    \caption{(Color online) Polarization resolved (a) experimental and (c) theoretical dependence of the transmitted intensity with the ellipticity $\theta$ of the incident beam, at high excitation power. The arrows mean forward and backward when changing $\theta$. 
    The dotted lines represent the expected behavior if $\alpha_{2}=0$. The degree of circular polarization corresponding the (a) and (c) is depicted in (b) and (d), respectively.}
    \label{figure3}
    \end{figure}

In order to understand the results presented in Figs.~\ref{figure1}-~\ref{figure2}, we have assumed that $\alpha_{2}$ is negligibly small, consistent with previous observations~\cite{Renucci2005,Ballarini2007}. However the actual value of $\alpha_{2}$ plays an important role when a high density of both $s_{z}=+1$ and $s_{z}=-1$ polaritons is simultaneously present in a given region of the sample. This is evidenced in Fig.~\ref{figure3}(a), where the transmitted intensity in a small region of $13\times13$~$\mu$m in the center of the spot, resolved into its $\sigma^{+}$ and $\sigma^{-}$ components, has been traced with respect to the ellipticity $\theta$ of the excitation beam in the conditions of Figs.~\ref{figure1}-~\ref{figure2}. Let us now analyze in detail the $\sigma^{-}$ curve [blue in Fig.\ref{figure3}(a)]. Here the argument will be equivalent for the $\sigma^{+}$ curve. When $\theta$ is close to $-\pi/2$, the excitation is purely $\sigma^{-}$. As the excitation density is significantly bigger than $D_{1}$, spin down polaritons lie on the ON state, the $s_{z}=-1$ polariton energy is renormalized to be in resonance with the excitation laser, and the transmitted intensity for the $\sigma^{-}$ polarization is high. On the contrary, spin up polaritons are in the OFF state. When the ellipticity of excitation is increased to $\theta=-0.21$~rad, the amount of spin up polaritons is big enough to induce a renormalization of the $s_{z}=+1$ branch such that the $\sigma^{+}$ component also jumps to the ON state. 
However, they do it to a value of transmitted intensity which is lower than that of the $\sigma^{-}$ polaritons at $\theta=-\pi/2$, and simultaneously, the $\sigma^{-}$ transmission decreases significantly. This is a direct consequence of the effective interaction between polaritons of the opposite spins, as also reported in~\cite{Paraiso2010}. At $k=0$, the energy of the $s_{z}=\pm1$ polaritons is given by:

\begin{equation}
E_{\pm}=E_{LP}(k=0)+\alpha_{1}\left|\Psi_{\pm}\right|^{2}+\alpha_{2}\left|\Psi_{\mp}\right|^{2}-\frac{i\hbar}{2\tau}
\label{eq:spininteraction}
\end{equation}

where $E_{LP}$ is the energy of the LPB in the absence of optical excitation, $\left|\Psi_{\pm}\right|^{2}$ is the $s_{z}=\pm1$ polariton density, and $\tau$ the polariton lifetime. Indeed, due to the non-zero value of $\alpha_{2}$, the presence of a large population of spin up polaritons for $\theta=-0.21\pi$~rad leads to the change in energy of the spin down polaritons given by Eq.~\ref{eq:spininteraction}, forcing them out of resonance with the excitation beam, and inducing the decrease of the $\sigma{^-}$ transmitted intensity. For the same reason, spin up polaritons do not reach the high transmitted intensity value expected if $\alpha_{2}=0$ , which is sketched in dashed lines in Fig.~\ref{figure3}(a). At a value of $\theta=0.22\pi$~rad, spin down polaritons fall to the OFF state, resulting in an increase of the transmitted intensity of the $\sigma^{+}$ polarization, as now the spin up polariton energy reaches the resonance. When changing the ellipticity in the backward direction, we observe the same phenomena, with slightly different thresholds for the jumps to the ON/OFF state ($\theta=-0.25\pi$ and $0.2\pi$~rad), due to the hysteretic behavior of our system.

Figure~\ref{figure3}(b) shows the degree of circular polarisation corresponding to Fig.~\ref{figure3}(a) [see Figs.~\ref{figure3}(c)-(d) for the corresponding simulations], where we observe that even with a non-zero value of $\alpha_{2}$, our system works as a very efficient polarization rectifier, with three possible output states: $\sigma^{-}$ polarized, linearly polarized, and $\sigma^{+}$ polarized.

    \begin{figure}[t]
        \centering
        \includegraphics[width=0.8\columnwidth]{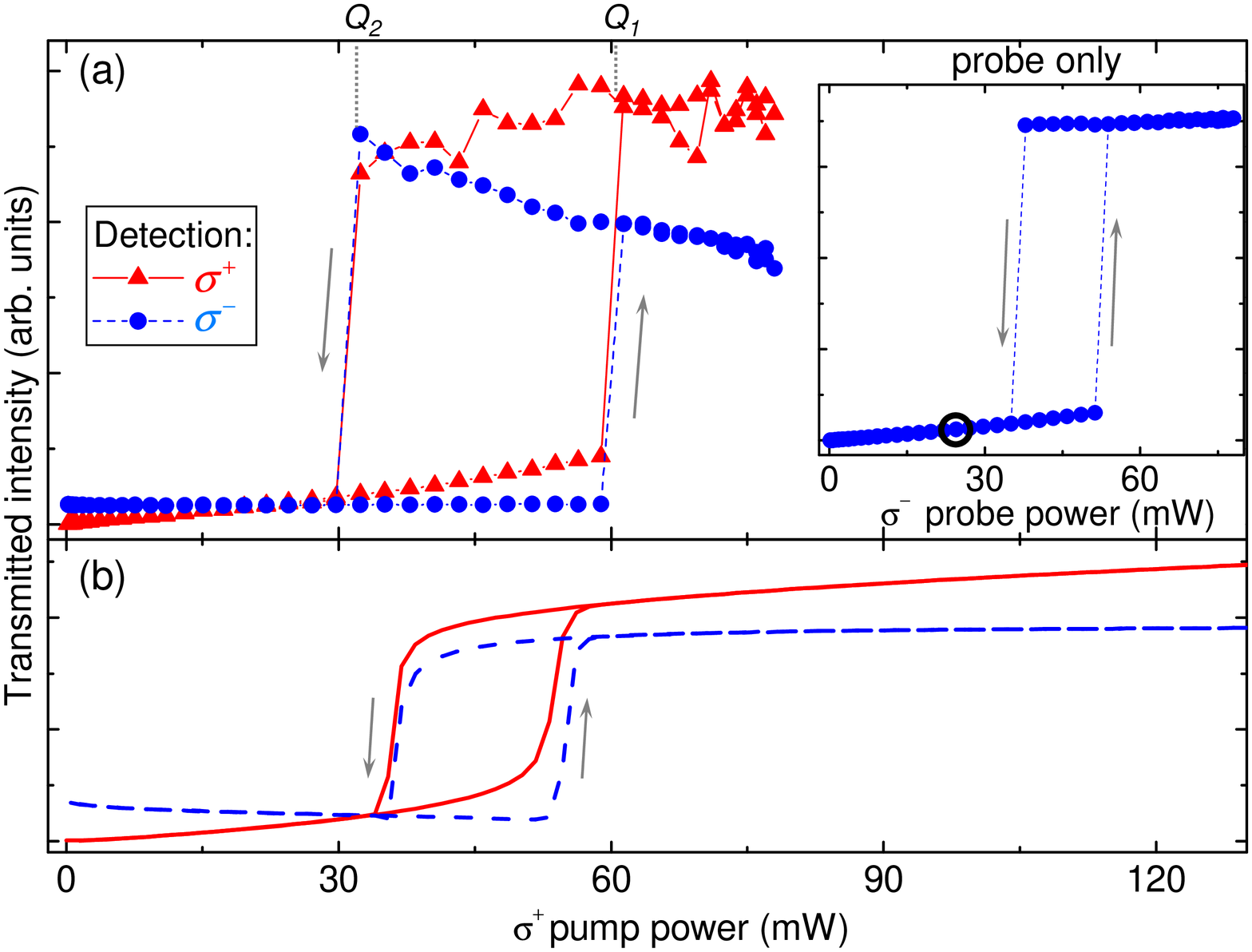}
    \caption{(Color online) (a) experimental polarization resolved transmission for a $\sigma^{+}$ pump whose power is varied in the presence of a $\sigma^{-}$ probe whose power is fixed (indicated by a circle in the inset, showing a power dependence of the probe alone). 
    }
    
    \label{figure4}
    \end{figure}

In Figs.~\ref{figure3}(a) and \ref{figure3}(b) we have evidenced the effects of a non-zero value of $\alpha_{2}$. In Fig.~\ref{figure4} we show an experiment that allows us to calculate its absolute value and sign. In this case we have a $\sigma^{-}$ probe beam, blue detuned by 0.2 meV from $E_{LPB}(k=0)$, whose power dependence is shown in the inset of Fig.~\ref{figure4}(a). We set the power of this probe to the value indicated by the circle, below the lowest point of the hysteresis region. Therefore, the probe beam alone would keep the system in the OFF state. Now we add a $\sigma^{+}$ pump beam, spatially overlapping the probe, whose density is varied. This is what is shown in Fig.~\ref{figure4}(a). The high density of $\sigma^{+}$ pump polaritons at point $Q_{1}$ (ON state) induces the renormalization of the $\sigma^{-}$ polariton energy (as given by Eq.~\ref{eq:spininteraction}), rendering $\sigma^{-}$ polaritons to the ON state. This is a clear indication that the sign of the effective $\alpha_{2}$ in our conditions is positive (repulsive interaction). A detailed fitting of Eq.~\ref{eq:spininteraction} to the data shown in Fig.~\ref{figure4}(a) is depicted in Fig.~\ref{figure4}(b). By performing similar fittings to analogous experiments for different probe powers, we obtain an effective $\alpha_{2}=+0.15\alpha_{1}$, which is the value employed in the simulations based on the spin-dependent Gross-Pitaevskii equation presented in Figs.~\ref{figure2}(b) and ~\ref{figure3}(c)-(d). This result seems to be at odds with recent theoretical~\cite{Wouters2007c,Glazov2009,Gippius2007,Ostatnicky2009} and experimental works performed in the optical parametric scattering regime~\cite{Renucci2005,Krizh2006b,Leyder2007b}, but in reality it is not, as the effective $\alpha_{2}$ may be influenced by a large fraction of dark incoherent excitons which contribute equally to the energy shift in $\sigma^{+}$ and $\sigma^{-}$ polarizations~\cite{Sarkar2010}. Our results are in agreement with recent reports under normal incidence pumping in a similar microcavity~\cite{Paraiso2010,Sarkar2010}. Further experiments are needed to measure the concentration of incoherent quasi-particles in our system, while clearly the incoherent fraction does not prevent spin switching and appearance of the spin rings.

In this work we have demonstrated the optical creation of spin ring domains in bi-stable semiconductor microcavities, whose size can be controlled down to the micrometer scale, well below the spot size. This arises from the strongly spin-dependent polariton-polariton interactions, an exceptional property of microcavity polaritons coming from their spin structure and strong light-matter coupling. 
Our results bring the polariton system closer to the implementation of integrated spin transistors~\cite{Johne2010} and logic devices~\cite{Liew2008} with very low thresholds and high potential operation speeds~\cite{Amo2010}.

We thank M.M. Glazov, D. Krizhanovskii, T. Ostatnick\'{y} and M. Vladimirova, for fruitful discussions. This work was supported by the \emph{Agence Nationale pour la Recherche} (GEMINI 07NANO 07043) and IFRAF. T.L. was supported by NCCR Quantum Photonics (NCCR QP), research instrument of the Swiss National Science Foundation (SNSF). A.K. acknowledges the support from the EU ETSF project n. 211956. A.B. is a member of the \emph{Institut Universitaire de France}.



\bibliography{articulosalberto}

\begin{thebibliography}{26}
\expandafter\ifx\csname natexlab\endcsname\relax\def\natexlab#1{#1}\fi
\expandafter\ifx\csname bibnamefont\endcsname\relax
  \def\bibnamefont#1{#1}\fi
\expandafter\ifx\csname bibfnamefont\endcsname\relax
  \def\bibfnamefont#1{#1}\fi
\expandafter\ifx\csname citenamefont\endcsname\relax
  \def\citenamefont#1{#1}\fi
\expandafter\ifx\csname url\endcsname\relax
  \def\url#1{\texttt{#1}}\fi
\expandafter\ifx\csname urlprefix\endcsname\relax\def\urlprefix{URL }\fi
\providecommand{\bibinfo}[2]{#2}
\providecommand{\eprint}[2][]{\url{#2}}

\bibitem[{\citenamefont{Eichten and Feinberg}(1981)}]{Eichten1981}
\bibinfo{author}{\bibfnamefont{E.}~\bibnamefont{Eichten}} \bibnamefont{and}
  \bibinfo{author}{\bibfnamefont{F.}~\bibnamefont{Feinberg}},
  \bibinfo{journal}{Phys. Rev. D} \textbf{\bibinfo{volume}{23}},
  \bibinfo{pages}{2724} (\bibinfo{year}{1981}).

\bibitem[{\citenamefont{Stenger et~al.}(1998)\citenamefont{Stenger, Inouye,
  Stamper-Kurn, Miesner, Chikkatur, and Ketterle}}]{Stenger1998}
\bibinfo{author}{\bibfnamefont{J.}~\bibnamefont{Stenger}},
  \bibinfo{author}{\bibfnamefont{S.}~\bibnamefont{Inouye}},
  \bibinfo{author}{\bibfnamefont{D.~M.} \bibnamefont{Stamper-Kurn}},
  \bibinfo{author}{\bibfnamefont{H.-J.} \bibnamefont{Miesner}},
  \bibinfo{author}{\bibfnamefont{A.~P.} \bibnamefont{Chikkatur}},
  \bibnamefont{and} \bibinfo{author}{\bibfnamefont{W.}~\bibnamefont{Ketterle}},
  \bibinfo{journal}{Nature} \textbf{\bibinfo{volume}{396}},
  \bibinfo{pages}{345} (\bibinfo{year}{1998}).

\bibitem[{\citenamefont{Gippius et~al.}(2007)\citenamefont{Gippius, Shelykh,
  Solnyshkov, Gavrilov, Rubo, Kavokin, Tikhodeev, and Malpuech}}]{Gippius2007}
\bibinfo{author}{\bibfnamefont{N.~A.} \bibnamefont{Gippius}},
  \bibinfo{author}{\bibfnamefont{I.~A.} \bibnamefont{Shelykh}},
  \bibinfo{author}{\bibfnamefont{D.~D.} \bibnamefont{Solnyshkov}},
  \bibinfo{author}{\bibfnamefont{S.~S.} \bibnamefont{Gavrilov}},
  \bibinfo{author}{\bibfnamefont{Y.~G.} \bibnamefont{Rubo}},
  \bibinfo{author}{\bibfnamefont{A.~V.} \bibnamefont{Kavokin}},
  \bibinfo{author}{\bibfnamefont{S.~G.} \bibnamefont{Tikhodeev}},
  \bibnamefont{and} \bibinfo{author}{\bibfnamefont{G.}~\bibnamefont{Malpuech}},
  \bibinfo{journal}{Phys. Rev. Lett.} \textbf{\bibinfo{volume}{98}},
  \bibinfo{pages}{236401} (\bibinfo{year}{2007}).

\bibitem[{\citenamefont{Para\"{i}so et~al.}(2010)\citenamefont{Para\"{i}so,
  Wouters, L\'{e}ger, Morier-Genoud, and Deveaud-Pl\'{e}dran}}]{Paraiso2010}
\bibinfo{author}{\bibfnamefont{T.~K.} \bibnamefont{Para\"{i}so}},
  \bibinfo{author}{\bibfnamefont{M.}~\bibnamefont{Wouters}},
  \bibinfo{author}{\bibfnamefont{Y.}~\bibnamefont{L\'{e}ger}},
  \bibinfo{author}{\bibfnamefont{F.}~\bibnamefont{Morier-Genoud}},
  \bibnamefont{and}
  \bibinfo{author}{\bibfnamefont{B.}~\bibnamefont{Deveaud-Pl\'{e}dran}},
  \bibinfo{journal}{Nature Mater. advanced online publication, DOI:
  10.1038/NMAT2787}  (\bibinfo{year}{2010}).

\bibitem[{\citenamefont{Zutic et~al.}(2004)\citenamefont{Zutic, Fabian, and
  Das~Sarma}}]{Zutic2004}
\bibinfo{author}{\bibfnamefont{I.}~\bibnamefont{Zutic}},
  \bibinfo{author}{\bibfnamefont{J.}~\bibnamefont{Fabian}}, \bibnamefont{and}
  \bibinfo{author}{\bibfnamefont{S.}~\bibnamefont{Das~Sarma}},
  \bibinfo{journal}{Rev. Mod. Phys.} \textbf{\bibinfo{volume}{\textbf{76}}},
  \bibinfo{pages}{323} (\bibinfo{year}{2004}).

\bibitem[{\citenamefont{Ono et~al.}(2002)\citenamefont{Ono, Austing, Tokura,
  and Tarucha}}]{Ono2002}
\bibinfo{author}{\bibfnamefont{K.}~\bibnamefont{Ono}},
  \bibinfo{author}{\bibfnamefont{D.~G.} \bibnamefont{Austing}},
  \bibinfo{author}{\bibfnamefont{Y.}~\bibnamefont{Tokura}}, \bibnamefont{and}
  \bibinfo{author}{\bibfnamefont{S.}~\bibnamefont{Tarucha}},
  \bibinfo{journal}{Science} \textbf{\bibinfo{volume}{\textbf{297}}},
  \bibinfo{pages}{1313} (\bibinfo{year}{2002}).

\bibitem[{\citenamefont{Vi\~{n}a et~al.}(1996)\citenamefont{Vi\~{n}a,
  Mu\~{n}oz, P\'{e}rez, Fernandez-Rossier, Tejedor, and Ploog}}]{Vina1996}
\bibinfo{author}{\bibfnamefont{L.}~\bibnamefont{Vi\~{n}a}},
  \bibinfo{author}{\bibfnamefont{L.}~\bibnamefont{Mu\~{n}oz}},
  \bibinfo{author}{\bibfnamefont{E.}~\bibnamefont{P\'{e}rez}},
  \bibinfo{author}{\bibfnamefont{J.}~\bibnamefont{Fernandez-Rossier}},
  \bibinfo{author}{\bibfnamefont{C.}~\bibnamefont{Tejedor}}, \bibnamefont{and}
  \bibinfo{author}{\bibfnamefont{K.}~\bibnamefont{Ploog}},
  \bibinfo{journal}{Phys. Rev. B} \textbf{\bibinfo{volume}{\textbf{54}}},
  \bibinfo{pages}{R8317} (\bibinfo{year}{1996}).

\bibitem[{\citenamefont{Nemec et~al.}(2005)\citenamefont{Nemec, Kerachian, van
  Driel, and Smirl}}]{Nemec2005}
\bibinfo{author}{\bibfnamefont{P.}~\bibnamefont{Nemec}},
  \bibinfo{author}{\bibfnamefont{Y.}~\bibnamefont{Kerachian}},
  \bibinfo{author}{\bibfnamefont{H.~M.} \bibnamefont{van Driel}},
  \bibnamefont{and} \bibinfo{author}{\bibfnamefont{A.~L.} \bibnamefont{Smirl}},
  \bibinfo{journal}{Phys. Rev. B} \textbf{\bibinfo{volume}{\textbf{72}}},
  \bibinfo{pages}{245202} (\bibinfo{year}{2005}).

\bibitem[{\citenamefont{Weisbuch et~al.}(1992)\citenamefont{Weisbuch, Nishioka,
  Ishikawa, and Arakawa}}]{Weisbuch1992}
\bibinfo{author}{\bibfnamefont{C.}~\bibnamefont{Weisbuch}},
  \bibinfo{author}{\bibfnamefont{M.}~\bibnamefont{Nishioka}},
  \bibinfo{author}{\bibfnamefont{A.}~\bibnamefont{Ishikawa}}, \bibnamefont{and}
  \bibinfo{author}{\bibfnamefont{Y.}~\bibnamefont{Arakawa}},
  \bibinfo{journal}{Phys. Rev. Lett.} \textbf{\bibinfo{volume}{\textbf{69}}},
  \bibinfo{pages}{3314} (\bibinfo{year}{1992}).

\bibitem[{\citenamefont{Wouters}(2007)}]{Wouters2007c}
\bibinfo{author}{\bibfnamefont{M.}~\bibnamefont{Wouters}},
  \bibinfo{journal}{Phys. Rev. B} \textbf{\bibinfo{volume}{76}},
  \bibinfo{pages}{045319} (\bibinfo{year}{2007}).

\bibitem[{\citenamefont{Glazov et~al.}(2009)\citenamefont{Glazov, Ouerdane,
  Pilozzi, Malpuech, Kavokin, and D'Andrea}}]{Glazov2009}
\bibinfo{author}{\bibfnamefont{M.~M.} \bibnamefont{Glazov}},
  \bibinfo{author}{\bibfnamefont{H.}~\bibnamefont{Ouerdane}},
  \bibinfo{author}{\bibfnamefont{L.}~\bibnamefont{Pilozzi}},
  \bibinfo{author}{\bibfnamefont{G.}~\bibnamefont{Malpuech}},
  \bibinfo{author}{\bibfnamefont{A.~V.} \bibnamefont{Kavokin}},
  \bibnamefont{and} \bibinfo{author}{\bibfnamefont{A.}~\bibnamefont{D'Andrea}},
  \bibinfo{journal}{Phys. Rev. B} \textbf{\bibinfo{volume}{80}},
  \bibinfo{pages}{155306} (\bibinfo{year}{2009}).

\bibitem[{\citenamefont{Ostatnicky et~al.}(2009)\citenamefont{Ostatnicky, Read,
  and Kavokin}}]{Ostatnicky2009}
\bibinfo{author}{\bibfnamefont{T.}~\bibnamefont{Ostatnicky}},
  \bibinfo{author}{\bibfnamefont{D.}~\bibnamefont{Read}}, \bibnamefont{and}
  \bibinfo{author}{\bibfnamefont{A.~V.} \bibnamefont{Kavokin}},
  \bibinfo{journal}{Phys. Rev. B} \textbf{\bibinfo{volume}{80}},
  \bibinfo{pages}{115328} (\bibinfo{year}{2009}).

\bibitem[{\citenamefont{Renucci et~al.}(2005)\citenamefont{Renucci, Amand,
  Marie, Senellart, Bloch, Sermage, and Kavokin}}]{Renucci2005}
\bibinfo{author}{\bibfnamefont{P.}~\bibnamefont{Renucci}},
  \bibinfo{author}{\bibfnamefont{T.}~\bibnamefont{Amand}},
  \bibinfo{author}{\bibfnamefont{X.}~\bibnamefont{Marie}},
  \bibinfo{author}{\bibfnamefont{P.}~\bibnamefont{Senellart}},
  \bibinfo{author}{\bibfnamefont{J.}~\bibnamefont{Bloch}},
  \bibinfo{author}{\bibfnamefont{B.}~\bibnamefont{Sermage}}, \bibnamefont{and}
  \bibinfo{author}{\bibfnamefont{K.~V.} \bibnamefont{Kavokin}},
  \bibinfo{journal}{Phys. Rev. B} \textbf{\bibinfo{volume}{\textbf{72}}},
  \bibinfo{pages}{075317} (\bibinfo{year}{2005}).

\bibitem[{\citenamefont{Baas et~al.}(2004)\citenamefont{Baas, Karr, Eleuch, and
  Giacobino}}]{Baas2004b}
\bibinfo{author}{\bibfnamefont{A.}~\bibnamefont{Baas}},
  \bibinfo{author}{\bibfnamefont{J.~P.} \bibnamefont{Karr}},
  \bibinfo{author}{\bibfnamefont{H.}~\bibnamefont{Eleuch}}, \bibnamefont{and}
  \bibinfo{author}{\bibfnamefont{E.}~\bibnamefont{Giacobino}},
  \bibinfo{journal}{Phys. Rev. A} \textbf{\bibinfo{volume}{69}},
  \bibinfo{pages}{023809} (\bibinfo{year}{2004}).

\bibitem[{\citenamefont{Shelykh et~al.}(2008)\citenamefont{Shelykh, Liew, and
  Kavokin}}]{Shelykh2008b}
\bibinfo{author}{\bibfnamefont{I.~A.} \bibnamefont{Shelykh}},
  \bibinfo{author}{\bibfnamefont{T.~C.~H.} \bibnamefont{Liew}},
  \bibnamefont{and} \bibinfo{author}{\bibfnamefont{A.~V.}
  \bibnamefont{Kavokin}}, \bibinfo{journal}{Phys. Rev. Lett.}
  \textbf{\bibinfo{volume}{100}}, \bibinfo{pages}{116401}
  (\bibinfo{year}{2008}).

\bibitem[{\citenamefont{Shelykh et~al.}(2006)\citenamefont{Shelykh, Rubo,
  Malpuech, Solnyshkov, and Kavokin}}]{Shelykh2006}
\bibinfo{author}{\bibfnamefont{I.~A.} \bibnamefont{Shelykh}},
  \bibinfo{author}{\bibfnamefont{Y.~G.} \bibnamefont{Rubo}},
  \bibinfo{author}{\bibfnamefont{G.}~\bibnamefont{Malpuech}},
  \bibinfo{author}{\bibfnamefont{D.~D.} \bibnamefont{Solnyshkov}},
  \bibnamefont{and} \bibinfo{author}{\bibfnamefont{A.}~\bibnamefont{Kavokin}},
  \bibinfo{journal}{Phys. Rev. Lett.} \textbf{\bibinfo{volume}{\textbf{97}}},
  \bibinfo{pages}{066402} (\bibinfo{year}{2006}).

\bibitem[{\citenamefont{Houdr\'{e} et~al.}(2000)\citenamefont{Houdr\'{e},
  Weisbuch, Stanley, Oesterle, and Ilegems}}]{Houdre2000}
\bibinfo{author}{\bibfnamefont{R.}~\bibnamefont{Houdr\'{e}}},
  \bibinfo{author}{\bibfnamefont{C.}~\bibnamefont{Weisbuch}},
  \bibinfo{author}{\bibfnamefont{R.~P.} \bibnamefont{Stanley}},
  \bibinfo{author}{\bibfnamefont{U.}~\bibnamefont{Oesterle}}, \bibnamefont{and}
  \bibinfo{author}{\bibfnamefont{M.}~\bibnamefont{Ilegems}},
  \bibinfo{journal}{Phys. Rev. B} \textbf{\bibinfo{volume}{\textbf{61}}},
  \bibinfo{pages}{R13333} (\bibinfo{year}{2000}).

\bibitem[{\citenamefont{Amo et~al.}(2009{\natexlab{a}})\citenamefont{Amo,
  Lefr\`{e}re, Pigeon, Adrados, Ciuti, Carusotto, Houdr\'{e}, Giacobino, and
  Bramati}}]{Amo2008}
\bibinfo{author}{\bibfnamefont{A.}~\bibnamefont{Amo}},
  \bibinfo{author}{\bibfnamefont{J.}~\bibnamefont{Lefr\`{e}re}},
  \bibinfo{author}{\bibfnamefont{S.}~\bibnamefont{Pigeon}},
  \bibinfo{author}{\bibfnamefont{C.}~\bibnamefont{Adrados}},
  \bibinfo{author}{\bibfnamefont{C.}~\bibnamefont{Ciuti}},
  \bibinfo{author}{\bibfnamefont{I.}~\bibnamefont{Carusotto}},
  \bibinfo{author}{\bibfnamefont{R.}~\bibnamefont{Houdr\'{e}}},
  \bibinfo{author}{\bibfnamefont{E.}~\bibnamefont{Giacobino}},
  \bibnamefont{and} \bibinfo{author}{\bibfnamefont{A.}~\bibnamefont{Bramati}},
  \bibinfo{journal}{Nature Phys.} \textbf{\bibinfo{volume}{\textbf{5}}},
  \bibinfo{pages}{805} (\bibinfo{year}{2009}{\natexlab{a}}).

\bibitem[{\citenamefont{Amo et~al.}(2009{\natexlab{b}})\citenamefont{Amo, Liew,
  Adrados, Giacobino, Kavokin, and Bramati}}]{Amo2009c}
\bibinfo{author}{\bibfnamefont{A.}~\bibnamefont{Amo}},
  \bibinfo{author}{\bibfnamefont{T.~C.~H.} \bibnamefont{Liew}},
  \bibinfo{author}{\bibfnamefont{C.}~\bibnamefont{Adrados}},
  \bibinfo{author}{\bibfnamefont{E.}~\bibnamefont{Giacobino}},
  \bibinfo{author}{\bibfnamefont{A.~V.} \bibnamefont{Kavokin}},
  \bibnamefont{and} \bibinfo{author}{\bibfnamefont{A.}~\bibnamefont{Bramati}},
  \bibinfo{journal}{Phys. Rev. B} \textbf{\bibinfo{volume}{80}},
  \bibinfo{pages}{165325} (\bibinfo{year}{2009}{\natexlab{b}}).

\bibitem[{\citenamefont{Ballarini et~al.}(2007)\citenamefont{Ballarini, Amo,
  Sanvitto, Vi\~{n}a, Skolnick, and Roberts}}]{Ballarini2007}
\bibinfo{author}{\bibfnamefont{D.}~\bibnamefont{Ballarini}},
  \bibinfo{author}{\bibfnamefont{A.}~\bibnamefont{Amo}},
  \bibinfo{author}{\bibfnamefont{D.}~\bibnamefont{Sanvitto}},
  \bibinfo{author}{\bibfnamefont{L.}~\bibnamefont{Vi\~{n}a}},
  \bibinfo{author}{\bibfnamefont{M.}~\bibnamefont{Skolnick}}, \bibnamefont{and}
  \bibinfo{author}{\bibfnamefont{J.}~\bibnamefont{Roberts}},
  \bibinfo{journal}{Appl. Phys. Lett.} \textbf{\bibinfo{volume}{\textbf{90}}},
  \bibinfo{pages}{201905} (\bibinfo{year}{2007}).

\bibitem[{\citenamefont{Krizhanovskii et~al.}(2006)\citenamefont{Krizhanovskii,
  Sanvitto, Shelykh, Glazov, Malpuech, Solnyshkov, Kavokin, Ceccarelli,
  Skolnick, and Roberts}}]{Krizh2006b}
\bibinfo{author}{\bibfnamefont{D.~N.} \bibnamefont{Krizhanovskii}},
  \bibinfo{author}{\bibfnamefont{D.}~\bibnamefont{Sanvitto}},
  \bibinfo{author}{\bibfnamefont{I.~A.} \bibnamefont{Shelykh}},
  \bibinfo{author}{\bibfnamefont{M.~M.} \bibnamefont{Glazov}},
  \bibinfo{author}{\bibfnamefont{G.}~\bibnamefont{Malpuech}},
  \bibinfo{author}{\bibfnamefont{D.~D.} \bibnamefont{Solnyshkov}},
  \bibinfo{author}{\bibfnamefont{A.}~\bibnamefont{Kavokin}},
  \bibinfo{author}{\bibfnamefont{S.}~\bibnamefont{Ceccarelli}},
  \bibinfo{author}{\bibfnamefont{M.~S.} \bibnamefont{Skolnick}},
  \bibnamefont{and} \bibinfo{author}{\bibfnamefont{J.~S.}
  \bibnamefont{Roberts}}, \bibinfo{journal}{Phys. Rev. B}
  \textbf{\bibinfo{volume}{\textbf{73}}}, \bibinfo{pages}{073303}
  (\bibinfo{year}{2006}).

\bibitem[{\citenamefont{Leyder et~al.}(2007)\citenamefont{Leyder, Liew,
  Kavokin, Shelykh, Romanelli, Karr, Giacobino, and Bramati}}]{Leyder2007b}
\bibinfo{author}{\bibfnamefont{C.}~\bibnamefont{Leyder}},
  \bibinfo{author}{\bibfnamefont{T.~C.~H.} \bibnamefont{Liew}},
  \bibinfo{author}{\bibfnamefont{A.~V.} \bibnamefont{Kavokin}},
  \bibinfo{author}{\bibfnamefont{I.~A.} \bibnamefont{Shelykh}},
  \bibinfo{author}{\bibfnamefont{M.}~\bibnamefont{Romanelli}},
  \bibinfo{author}{\bibfnamefont{J.~P.} \bibnamefont{Karr}},
  \bibinfo{author}{\bibfnamefont{E.}~\bibnamefont{Giacobino}},
  \bibnamefont{and} \bibinfo{author}{\bibfnamefont{A.}~\bibnamefont{Bramati}},
  \bibinfo{journal}{Phys. Rev. Lett.} \textbf{\bibinfo{volume}{\textbf{99}}},
  \bibinfo{pages}{196402} (\bibinfo{year}{2007}).

\bibitem[{\citenamefont{Sarkar et~al.}(2010)\citenamefont{Sarkar, Gavrilov,
  Sich, Quilter, Bradley, Gippius, Guda, Kulakovskii, Skolnick, and
  Krizhanovskii}}]{Sarkar2010}
\bibinfo{author}{\bibfnamefont{D.}~\bibnamefont{Sarkar}},
  \bibinfo{author}{\bibfnamefont{S.~S.} \bibnamefont{Gavrilov}},
  \bibinfo{author}{\bibfnamefont{M.}~\bibnamefont{Sich}},
  \bibinfo{author}{\bibfnamefont{J.~H.} \bibnamefont{Quilter}},
  \bibinfo{author}{\bibfnamefont{R.~A.} \bibnamefont{Bradley}},
  \bibinfo{author}{\bibfnamefont{N.~A.} \bibnamefont{Gippius}},
  \bibinfo{author}{\bibfnamefont{K.}~\bibnamefont{Guda}},
  \bibinfo{author}{\bibfnamefont{V.~D.} \bibnamefont{Kulakovskii}},
  \bibinfo{author}{\bibfnamefont{M.~S.} \bibnamefont{Skolnick}},
  \bibnamefont{and} \bibinfo{author}{\bibfnamefont{D.~N.}
  \bibnamefont{Krizhanovskii}}, \bibinfo{journal}{(submitted)}
  (\bibinfo{year}{2010}).

\bibitem[{\citenamefont{Johne et~al.}(2010)\citenamefont{Johne, Shelykh,
  Solnyshkov, and Malpuech}}]{Johne2010}
\bibinfo{author}{\bibfnamefont{R.}~\bibnamefont{Johne}},
  \bibinfo{author}{\bibfnamefont{I.~A.} \bibnamefont{Shelykh}},
  \bibinfo{author}{\bibfnamefont{D.~D.} \bibnamefont{Solnyshkov}},
  \bibnamefont{and} \bibinfo{author}{\bibfnamefont{G.}~\bibnamefont{Malpuech}},
  \bibinfo{journal}{Phys. Rev. B} \textbf{\bibinfo{volume}{81}},
  \bibinfo{pages}{125327} (\bibinfo{year}{2010}).

\bibitem[{\citenamefont{Liew et~al.}(2008)\citenamefont{Liew, Kavokin, and
  Shelykh}}]{Liew2008}
\bibinfo{author}{\bibfnamefont{T.~C.~H.} \bibnamefont{Liew}},
  \bibinfo{author}{\bibfnamefont{A.~V.} \bibnamefont{Kavokin}},
  \bibnamefont{and} \bibinfo{author}{\bibfnamefont{I.~A.}
  \bibnamefont{Shelykh}}, \bibinfo{journal}{Phys. Rev. Lett.}
  \textbf{\bibinfo{volume}{101}}, \bibinfo{pages}{016402}
  (\bibinfo{year}{2008}).

\bibitem[{\citenamefont{Amo et~al.}(2010)\citenamefont{Amo, Liew, Adrados,
  Houdre, Giacobino, Kavokin, and Bramati}}]{Amo2010}
\bibinfo{author}{\bibfnamefont{A.}~\bibnamefont{Amo}},
  \bibinfo{author}{\bibfnamefont{T.~C.~H.} \bibnamefont{Liew}},
  \bibinfo{author}{\bibfnamefont{C.}~\bibnamefont{Adrados}},
  \bibinfo{author}{\bibfnamefont{R.}~\bibnamefont{Houdre}},
  \bibinfo{author}{\bibfnamefont{E.}~\bibnamefont{Giacobino}},
  \bibinfo{author}{\bibfnamefont{A.~V.} \bibnamefont{Kavokin}},
  \bibnamefont{and} \bibinfo{author}{\bibfnamefont{A.}~\bibnamefont{Bramati}},
  \bibinfo{journal}{Nature Phot.} \textbf{\bibinfo{volume}{4}},
  \bibinfo{pages}{361} (\bibinfo{year}{2010}).

\end{thebibliography}

\end{document}